\documentclass[english]{article}
\usepackage[T1]{fontenc}
\usepackage[latin9]{inputenc}
\usepackage{graphicx}
\usepackage{babel}
\usepackage{amsmath}
\usepackage{multicol}
\makeatletter

\newcommand{\fig}[1]{Figure (\ref{#1})}
\newcommand{\eq}[1]{Eq. (\ref{#1})}
\newcommand{\eqs}[1]{Eqs. (\ref{#1})}

\newcommand{\la}[1]{ \label{#1}}

\renewcommand{\b}{\beta}

\newcommand{\bsubs}{\begin{subequations}}
\newcommand{\esubs}{\end{subequations}}

\newcommand{\be}{\begin{equation}}
\newcommand{\ee}{\end{equation}}

\newcommand{\bea}{\begin{eqnarray}}
\newcommand{\eea}{\end{eqnarray}}

\begin{document}

\title{More is the Same;\\
 Phase Transitions and Mean Field Theories\\
}
\author{ Leo P. Kadanoff\\
The James Franck Institute\\
The University of Chicago\\
email: leop@UChicago.edu}

\maketitle
\begin{abstract}

This paper is the first in a series that will look at the theory of phase transitions from the perspectives of physics and the philosophy of science.   The series will consider a group of related concepts derived from condensed matter and statistical physics.  The key technical ideas go under the names of  ``singularity'', ``order parameter'', ``mean field theory'',   ``variational method'', ``correlation length'', ``universality class'', ``scale changes'', and ``renormalization''.   The first four of these will be considered here.

In a less technical vein, the question here is how can matter, ordinary matter, support a diversity of forms.   We see this diversity each time we observe ice in contact with liquid water or see water vapor (steam) come up from a pot of heated water. Different phases can be qualitatively different in that walking on ice is well within human capacity, but walking on liquid water is proverbially forbidden to ordinary humans.  These differences have been apparent to humankind for millennia, but only brought within the domain of  scientific understanding since the 1880s.   

A phase transition is a change from one behavior to another.  A first order phase transition involves a discontinuous jump in some statistical variable.  The discontinuous property is called the order parameter. Each phase transition has its own order parameter.  The possible order parameters range over a tremendous variety of physical properties.  These properties include the density of a liquid-gas transition, the magnetization in a ferromagnet, the size of a connected cluster in a percolation transition, and a condensate wave function in a superfluid or superconductor. A continuous transition occurs when the discontinuity in the jump approaches zero.   This article is about  statistical mechanics and the development of   mean field theory as a basis for a partial understanding of phase transition phenomena.   

Much of the material in this review was first prepared for the Royal Netherlands Academy of Arts and Sciences in 2006.   It has appeared in draft form on the authors' web site\cite{LPKwebsite} since then.  

The title of this article is a {\em hommage} to Philip Anderson and his essay ``More is Different,''\cite{PWA} \cite{reviewPWA} which describes how new concepts, not applicable in ordinary classical or quantum mechanics,    can arise from the consideration of aggregates of large numbers of particles. Since phase transitions {\em only} occur in systems with an infinite number of degrees of freedom, such transitions are a prime example of Anderson's thesis.

\end{abstract}
\tableofcontents

\newpage{ }
\section{Introduction}
\begin{description}
  \item[]    The Universe was brought into being in a less than fully formed state, but was gifted with the capacity to transform itself from unformed matter into a truly marvelous array of physical structures ...
  \item[ Saint Augustine of Hippo  (354-430).         \	  Translation by Howard J. van Till\cite{vanTill}]
\end{description}

\subsection{Phases}

Matter exists in different thermodynamic ``phases'', which are different states of aggregation with qualitatively different properties. These phases provoked studies that are instructive to the history of science.  The phases themselves are interesting to modern physics, and are provocative to modern philosophy.   For example, the philosopher might  wish to note that, strictly speaking,   no phase transition can ever occur in a finite system.  Thus, in some sense, phase transitions are   not exactly embedded in the finite world but, rather, are products of the human imagination.  

Condensed matter physics is a branch of physics dealing with the properties of the bulk matter around us.  This matter arranges itself into structures that are amazingly diverse and beautiful.  \fig{iceberg} illustrates three of the many thermodynamic phases formed by water.   The solid iceberg sits in contact with liquid water and with the water vapor in the air above.   Water has many different solid phases.  Other fluids form liquid crystals, in which we can see  macroscopic manifestations of the shapes of the molecules forming the crystals. The alignment of atomic spins or electronic orbits can produce diverse magnetic materials, including ferromagnets, with their substantial magnetic fields, and also many other more subtle forms of magnetic ordering. Our economic infrastructure is, in large measure,  based upon the various phase-dependent capabilities of materials to carry electrical currents:  from the refusal of insulators, to the flexibility of semiconductors,  to the substantial carrying capacity of conductors, to the weird resistance-free behavior of superconductors.  I could go on and on.  The point is that humankind has, in part, understood these different manifestations of matter, manifestations that go under the name ``thermodynamic phases''. Scientific work has produced at least a partial understanding of how the diffferent phases change into one another: a process called phase transitions.    This article is a brief description of the ideas contained in the science of such things.

\begin{figure}
\begin{multicols}{2}
\includegraphics[height=7cm ]{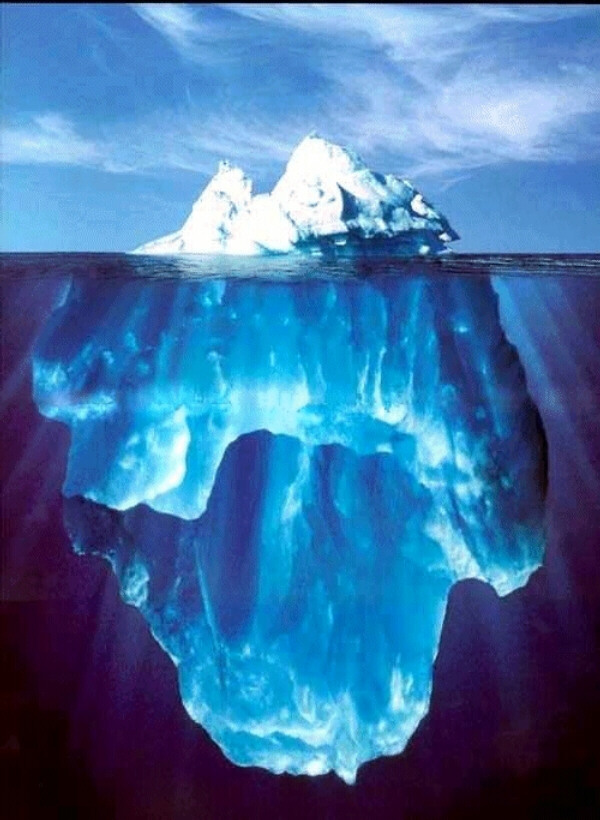}
\caption{Iceberg floating in the sea.  This picture is intended to illustrate different phases of water. The sea is liquid water, which is then in contact with solid water in the form of ice.   In the air above, breezes blow clouds through the air, which  contains water vapor in contact with both the solid and the liquid forms of water.  The change from one form to the next is termed a {\em phase transition.}  }
\la{iceberg}
\end{multicols}
\end{figure}

\subsection{A phase diagram}

Paul Ehrenfest\cite{Ehrenfest,Daugherty} attempted to  classify  the different kinds of phase transitions.  All phase transitions involve a sudden qualitative change in material behavior  or precipitous changes in some thermodynamic quantities. Mathematicians describe these sudden changes as {\em singularities}. Ehrenfest classified phase transitions by sorting the different  kinds of thermodynamic quantities that undergo discontinuous jumps.   The modern classification scheme for  singularities in phase transitions is both simpler than Ehrenfest's original scheme and more complex.  We note the existence of more complex mathematical singularities than Ehrenfest's simple jumps\footnote{Ehrenfest gave numbers to the different kinds of phase transitions: first, second, third, ....  However, it turns out that there are are too many different kinds of continuous transitions for a scheme like Ehrenfest's   to work.}. But, we define only two fundamentally different kinds of phase transitions: { \em First order phase transitions} are ones in which basic thermodynamic quantities like the number of particles per unit volume or the magnetization show a sudden jump as a function of temperature or pressure or other thermodynamic characteristic.  {\em Continuous  transitions}  are those in which some sudden changes occur, but these changes are more gentle than a discontinuous jump in the basic variables.      The positions in phase diagrams at which we see continuous transitions are called {\em critical points}.   
  
    \fig{ferro} shows a phase diagram for a simplified ferromagnetic system. The ferromagnet is characterized by the possibility of having a strong magnetization caused by the alignment of atomic spins within the material.    We imagine that our ferromagnet has one special crystal axis, the ``easy axis''. The magnetization of this material is a vector forced to point  along    or against the direction set by this easy axis. This behavior is ensured in part by the internal structure of the material and in part by insisting that, when the material  is placed in a magnetic field,  that field also has an alignment set by the easy axis. The basic variables defining the state of the system are  the magnetic field and the temperature.  We describe what is happening by looking at the magnetization.  The magnitude of the magnetization measures the extent to which the spins in the system are lined up with each other.   Its sign describes the direction of the allignment.   If the temperature is sufficiently low, the system has a non-zero magnetization even at zero magnetic field.   At  these lower temperatures, the zero-field magnetization has two possible values, for the two possible directions in which the spins may align themselves.    The heavy line in the phase diagram, \fig{ferro}, is the locus of points at which this spontaneous magnetization is non-zero.  As one crosses this line, there is a discontinuous jump in the magnetization, which maintains its magnitude but reverses its direction.  This jump is a first order phase transition.  Typically, this jump decreases in size as the temperature gets higher until, at some critical point, the jump goes continuously to zero.   This point is then the position of a continuous phase transition.

\begin{figure}
\begin{centering}
\includegraphics[height=8cm ]{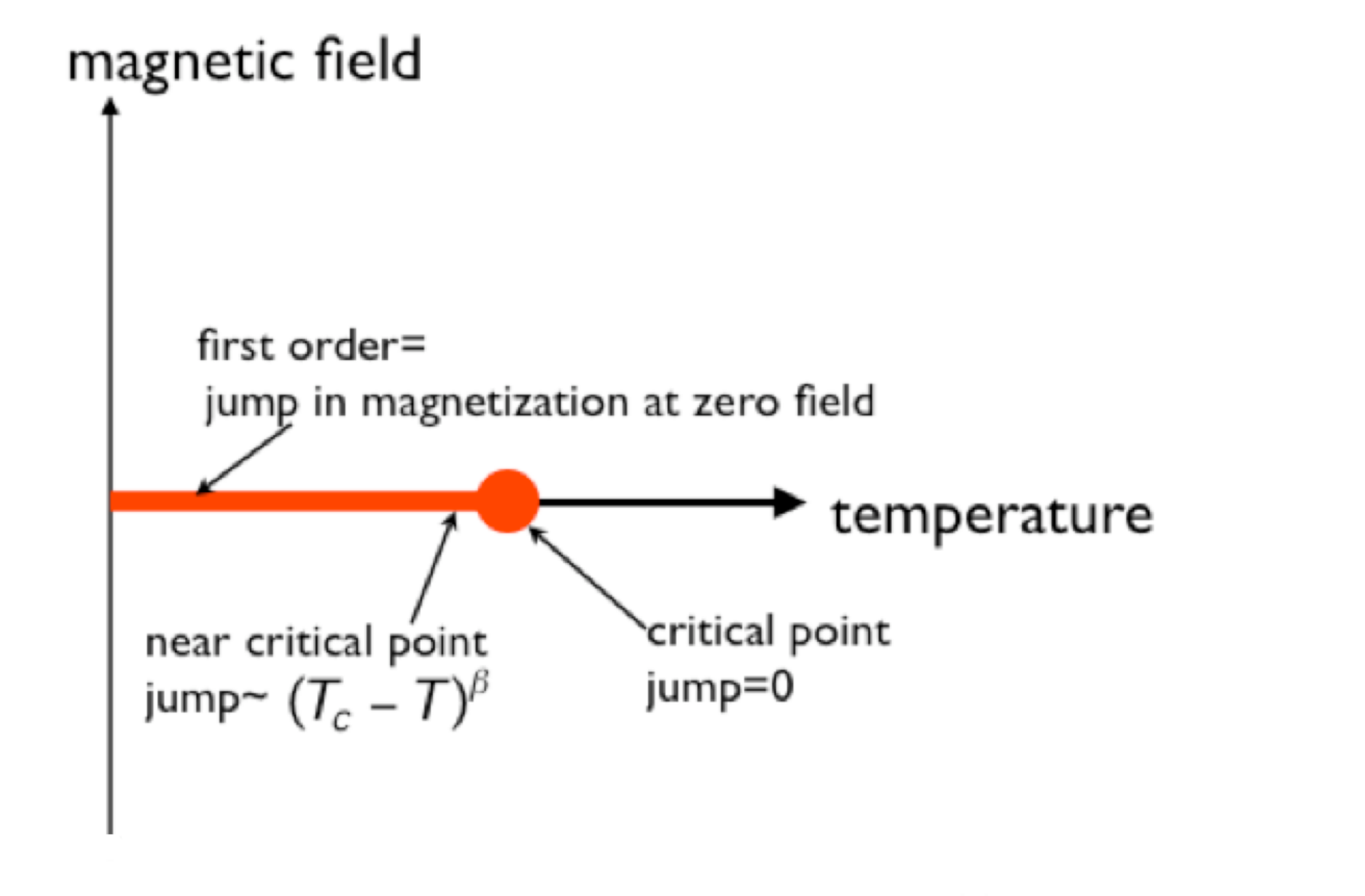}
\end{centering}
\caption{Phase diagram for ferromagnet.  The heavy line is at zero magnetic field. Along this line there is a first order phase transition. As the magnetic field approaches zero through positive values, the magnetization approaches a non-zero positive value on the line. As the magnetic field approaches zero through negative values, the magnetization approaches a non-zero negative value on the line.  As the field passes through zero on the line, the magnetization goes through a jump by changing sign. The heavy dot at the end of the line indicates a critical point.  At zero magnetic field and temperatures above this critical point the magnetization is strictly zero.  As one passes below the critical point, the magnetization continuously increases in magnitude from its zero value while maintaining whatever  sign it happened to have at the critical point.    }
\la{ferro}
\end{figure}

\subsection{The beginning}
The phase diagram and the Ehrenfest classification are far from the beginning of the modern phase transition story.  One beginning is with J. Willard Gibbs, who both defined  modern statistical mechanics and extensively studied thermodynamics, including the thermodynamics  of  phase transitions\cite{Gibbs-E,Gibbs-S}. The definition of statistical mechanics starts from the system's configurations.  It then gives the probability of a particular configuration, labeled by $c$, as
\begin{equation}
\rho(c) = e^{- H(c)/T }/Z
\la{rho}
\ee		
where $T$ is the  temperature measured  in energy units\footnote{More conventionally, one would write instead of $T$ the product $kT$, where $k$ is the Boltzmann constant.} and $H$ is the Hamiltonian, so that $H(c)$ is the energy of configuration $c$.  The partition function, $Z$, is given as the sum over all configurations  
\be
Z= e^{- F/T } = \sum_c  e^{- H(c)/T },
\la{Z}
\ee	                 
which, then, also defines the free energy, F.

\subsection{The Ising model}
We start with a model for a system that can potentially show ferromagnetic behavior.  The simplest model, in extensive use today\footnote{Much of the historical material in this work is taken from the excellent book on critical phenomena  by Cyril Domb\cite{Domb}.},  is called the Ising model, after the physicist Ernest Ising\cite{Ising}, who invented it in conjunction with his adviser Wilhelm Lenz\cite{Lenz}.   Real ferromagnets involve atomic spins placed upon a lattice. The elucidation of their properties requires a difficult study via the band theory of solids. The Ising model is a shortcut that catches the main qualitative features of the ferromagnet.  It puts a spin variable, $\sigma$,  upon each site-labeled by $\bf{r} $ of a simple lattice.  (See \fig{lattice}.)  Each spin variable takes on values plus or minus one to represent the possible directions that might be taken by a particular component of a real spin upon a real atom.   So the system has been reduced to a set of  variables, $\sigma_{\bf{r}}$,  each taking on one of two possible values.\footnote{Ludwig Boltzmann would have felt particularly at home with this particular example  of a statistical system, since he preferred discrete analysis of probabilities to continuous ones.  See, for example, the article of Giovanni Gallavotti\cite{GG} in the Boltzmann centennial volume\cite{LB}.}     The sum over configurations is a sum over these possible values of all the different spin variables at all the lattice sites.   The Hamiltonian for the system is the simplest representation of the fact that neighboring spins interact with a dimensionless coupling strength, $K$, and a dimensionless coupling to an external magnetic field, $h$.  The Hamiltonian is given by
\be
-H/T=K  \sum_{nn} \sigma_r \sigma_s+ h \sum_r \sigma_r
\la{Ising}
\ee         
where the first sum is over all pairs of nearest neighboring sites, and the second is over all sites.  The actual coupling between neighboring spins, with dimensions of an energy, is often called $J$. Then $K=-J/ T.$  In turn, $h$ is proportional to the magnetic moment of the given spin times the applied magnetic field, all divided by the temperature.
\begin{figure}
\begin{multicols}{2}
\includegraphics[height=5cm ]{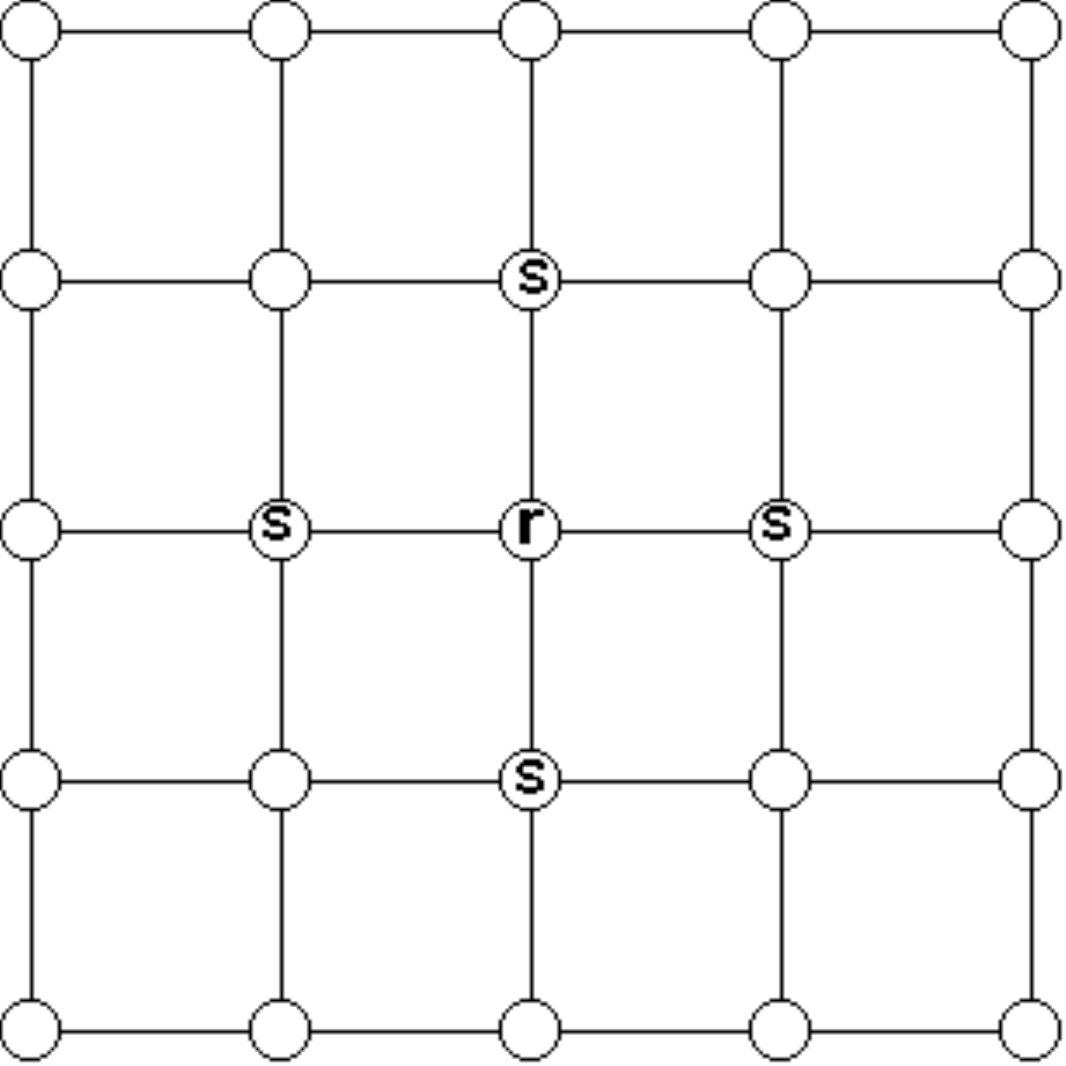}
\caption{
Lattice for two dimensional Ising model.  The spins are in the circles.  The couplings, $K$, are the lines.  A particular site is labeled with an ``r''.  Its nearest neighbors are shown with an ``s''.    }
\la{lattice}
\end{multicols}
\end{figure} 

\fig{ferro} is the phase diagram of the Ising model.  The $x$ axis is then $1/K$ and the $y$ axis is $h$.   This phase diagram  applies when the lattice is infinite in two or more dimensions.      

\subsection{More is the same; infinitely more is different}

In discussing phase transitions, we must note a point that is  fundamental to condensed matter physics.  In the words of Anderson\cite{PWA}, {\em more is different}.   The properties of systems containing infinitely many particles are qualitatively different from those of finite systems.
In particular, phase transitions cannot occur in any finite system; they are solely a property of infinite systems.   

To see this we  follow Ehrenfest and Gibbs and define a phase transition as a point of singularity, i.e. sudden change.   Let us see what this definition implies about any finite Ising model, one containing a finite number of spin variables.  A phase transition occurs at points in the phase diagram where the free energy is a singular function of  the thermodynamic variables within it.     The partition function and the free energy are defined in \eq{Z}.  The former  is the sum of the exponential of $- H/T$ over all possible  configurations.  Such a sum of a finite number of exponentials is necessarily a positive quantity and one that is regular, i.e. not singular, for any finite value of $K$ or $h$.   Such a nonsingular quantity can have no sudden changes. Taking the logarithm of a positive quantity introduces no singularities, nor does division by a finite number, the temperature.   Hence, we can conclude that,  for the finite Ising model, the free energy is a non-singular function of $K$ and $h$ for all finite, real values of these parameters.   By this argument the  Ising model, as we have described it, can have no phase transitions.   Further,   there cannot be any phase transition in any finite system described by the Ising model or indeed any statistical system with everything in it being finite.  

But phase transitions certainly do occur.  And the Ising model is a pretty decent representation of a ferromagnet.  Where is the hole in our argument?   The hole can be seen by using continuity arguments, and equally well from modern numerical studies that show the nature of the ``discontinuous jump''.  A finite sum of exponentials cannot possibly give a singularity, but an infinite sum can indeed do so.    Arthur Wightman\cite{Wightman} has emphasized that  Gibbs could certainly have known that phase transition are properties of infinite systems.    However, Gibbs' book on statistical mechanics\cite{Gibbs-E} never did anything quite as specific as discuss phase transitions.  According to George E. Uhlenbeck  
\cite{Uhlenbeck},  Hendrick A. Kramers was the first to point out that the sharp singularity of a phase transition could only occur in a system with  some infinity built in. The usual infinity is an infinite number of lattice sites or particles.  Apparently the point remained contentious as late as 1937.  I quote E. G. D.  Cohen's description of material contained in another work  of Uhlenbeck\cite{Cohen}:  ``Apparently the audience at this Van der Waals memorial  meeting in 1937, could not agree on the above question, whether the partition function could or could not explain a sharp phase transition. So the chairman of the session, Kramers, put it to a vote."

The point about phase transitions being a property of infinite systems is an important clue to the correct characterization of these transitions.  If these systems  must be infinite, then the phase transition is likely to be characterized by long-ranged order within the system, and that order must be important in the far reaches of the system. Any theory of phase transitions that does not include ordering at infinity is likely to be inadequate. Many years after Gibbs,  David Ruelle\cite{Ruelle} put together a mathematically precise theory of phase transitions, and of course it centered upon the interaction of far-away boundaries and ordering\footnote{This paper emphasizes that phase transitions in materials are connected with singularities created by the system's infinite spacial extension.   Other possible sources of singularities exist. On possibility, important in particle physics, is an {\em ultraviolet divergence}, that is an infinite number of degrees of freedom appearing within a finite volume.  Another is an interaction which is infinitely strong.  A third possibility is that the definition of the statistical average itself includes an infinity.  This last possibility is realized in the microcanonical ensemble, which includes an infinitely sharp peak in energy\cite{Dunkel}.}.

\begin{figure}
\begin{centering}
\includegraphics[width=7cm ]{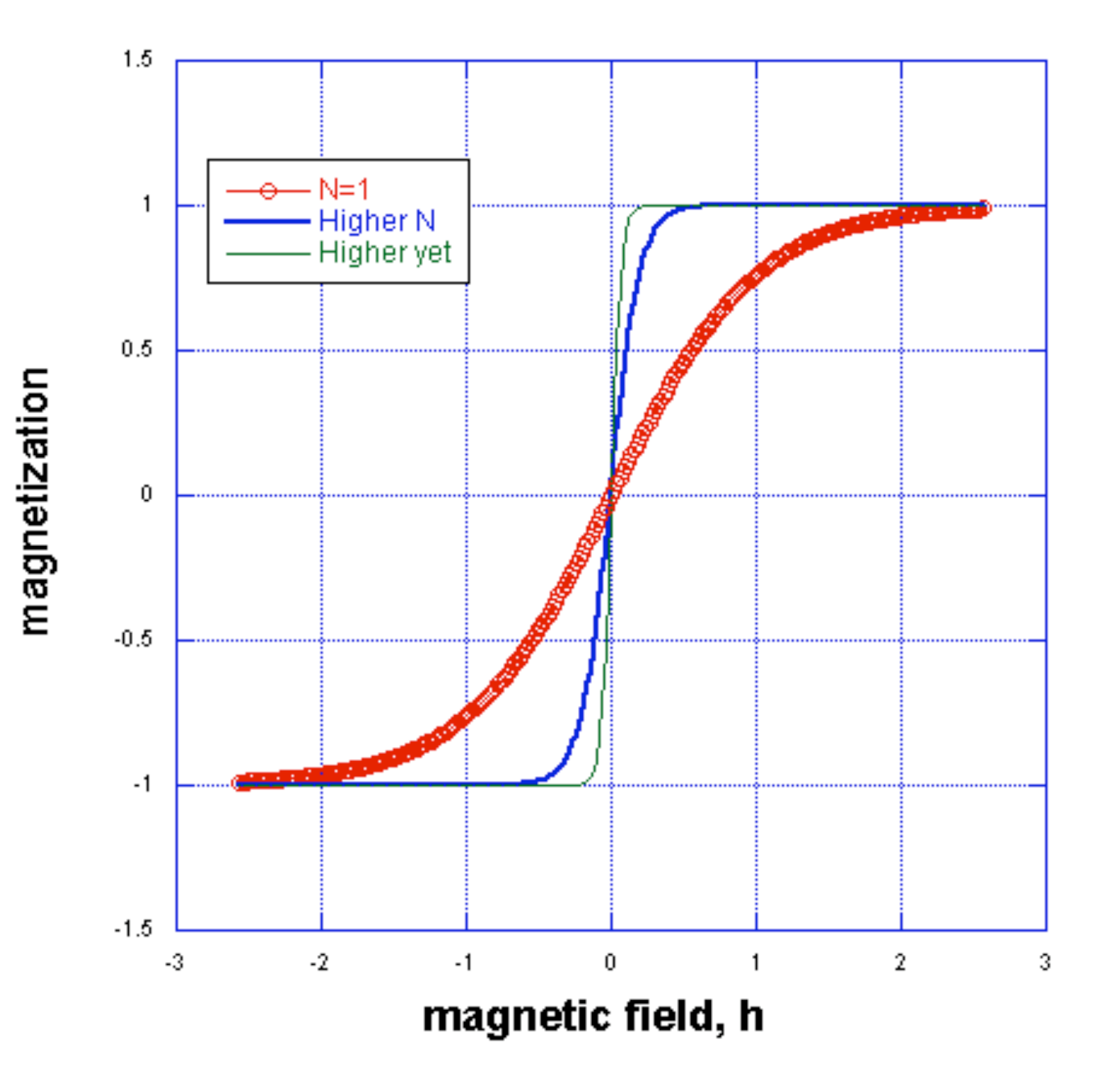}
\end{centering}
\caption{Magnetization, $<\sigma>$, versus magnetic field, $h$.  The $N=1$ curve is a plot of  \eq{av} showing magnetization versus $h$ for one spin.   The other two curves are cartoon views showing what would happen for temperatures below the critical temperature if one increased the number of particles beyond one, but kept the number reasonably small.  }
\la{magvsh}
\end{figure}

Numerical studies back up this conclusion that phase transitions only occur in infinitely large systems.  If we look at the magnetization in a series of different systems varying in their numbers of lattice sites, we shall see a characteristic variation of the sharpness of the jump with the number of sites in the lattice.  Imagine varying the magnetic field at fixed temperature, as in \fig{magvsh}.  For systems with fewer lattice sites the magnetization will vary slowly and continuously through zero as the field passes through zero.   As the number of lattice sites gets larger the variation in the magnetization will get steeper, until at a very large number of sites the transition from positive values of $<\sigma> $ to negative ones will become so steep that the casual observer might say that it has occurred suddenly.    The astute observer will look more closely, see that there is a very steep rise, and perhaps conclude that the discontinuous jump only occurs in the infinite system. 

\subsection{Ising-model results}

But back to the model of  \eq{Ising}.  Ising calculated the free energy for the one-dimensional version of his model.  In this case, the spins are distributed along a line, either finite or infinite in length.  In either case,  Ising saw no phase transition in this one-dimensional model.    

Lev Landau\cite{1d} later argued that there was no phase transition in any finite-temperature system in one dimension.   So, by that argument, the first situation in which we might expect to find a phase transition is in an infinite two-dimensional system.    
 
A portion of this discussion was brought to a culmination by Rudolf Peierls\cite{Peierls} and Robert Griffiths\cite{Griffiths},  who proved that phase transitions exist within statistical mechanics by showing that  the Ising model has a zero-field magnetization in two  dimensions and a sufficiently low temperature.  Since the zero-field magnetization must be zero at sufficiently high temperatures,  a qualitative change in behavior must occur at some intermediate temperature.    

By this argument whenever a material shows one behavior in one region of a phase diagram, and another, qualitatively different, behavior in a different region then somewhere in between there must be a phase transition.  The fact that you can walk on ice but not on liquid water strongly suggests that in between one should find a phase transition.   Thus, we can infer that the melting of ice is indeed a phase transition.    


\subsection{Why study this model, or any model?} 
These results enable us to see at once the answer to the question ``why study a simplified model like the Ising model?''  The answer is that the simplicity enables one to formulate and test important qualitative questions such as ``in what situations might it be possible to have a phase transition?''   The answer to this  question is not obvious and can  be best approached through the simplest model that does display the phase transition.  Many such deep questions can be studied by using highly simplified models\cite{RB}.  This strategy is made rewarding by a characteristic of physical systems called ``universality'', in that many systems may show the very same qualitative features, and sometimes even the same quantitative ones.  To study a given qualitative feature, it often pays to look for the simplest possible example. 

This brings us to the next issue:  ``Are phase transitions real?'' That question bears upon the source of physical concepts.  Since a phase transition only happens in an infinite system, we cannot say that any phase transitions actually occur in the finite objects that appear in our world.   Instead, we must conclude that phase transitions and the definitions of different  thermodynamic phases are the result of a process of extrapolating the ``real'' behavior of a theory of large bodies, to its infinite conclusion\cite{Batterman,Berry}. Indeed, Nature gives us no pure thermodynamic phases but only real objects displaying their own complex and messy behavior.    This extrapolation and simplifying process necessary to define thermodynamic phases suggests that at least this part of theoretical physics is not a simple result of the direct examination of Nature, but rather it is a result of the  human imagination applied to an extrapolation of that examination.    In this way, the Ising model helps us see more clearly what our imagination has produced.  No addition of bells, whistles, or additional complicating features could, for this purpose, improve upon the bare simplicity of the Ising model.  The point is not to give an accurate description of a particular ferromagnet, but rather to give a barebones description from which one can infer general features of a ferromagnet.  

\section{More is the same: Mean Field Theory}
This section is about the concept of a mean field (or effective field),   which forms the basis of much of modern condensed matter physics and also of particle physics. 

We look at mean field theory because it helps remove a big hole in our previous discussion.  So far, we said that sometimes an infinite statistical system has a phase transition, and that transition involves a discontinuous jump in a quantity we call the order parameter.   But we have given no indication of how big the jump might be, nor of how the system might produce it.   Mean field theory provides a partial, and partially imprecise, answer to that question.   Of course, ``partially imprecise'' means that the answer is partially right.  That is the part we are after.

We begin with the statistical mechanics of one spin in a magnetic field. Then, we extend this one-spin discussion to describe how many spins work together to produce ferromagnetism.

\subsection{One spin}
A single spin in a magnetic field can be described by a simplified version of the Ising Hamiltonian \eq{Ising}.  Throw away the sums, throw away the coupling among spins and you are left with a Hamiltonian given by
\be
- H/T = h  \sigma
\la{one}
\ee 
As before, $\sigma$  is a component of the spin in the direction of the magnetic field, which is defined to point parallel to some crystal axis, the ``easy axis''.   The field has the dimensionless representation $h$.  The quantum variable takes on two values $\pm 1$, so that  the basic statistical mechanics  of \eq{rho} gives the probabilities of finding the spins with the two different values as 
$$
\rho(+1)=e^{h}/z  \text{~ and ~} \rho(-1)=e^{-h}/z  \text{~ with ~}   z=e^h+e^{-h}=2 \cosh h
$$
so that the average value of the spin is 
\be
<\sigma> =e^{h}/z-e^{-h}/z  = \tanh h
\la{av}
\ee
Thus, as we might expect, the statistical mechanics formulation makes the average magnetization, $<\sigma> $, increase smoothly from minus one to zero to one as the component of the magnetic field along the easy axis increases smoothly from minus infinity to zero to infinity. This behavior is depicted in \fig{magvsh}.

\subsection{Curie-Weiss many spins; mean fields}
The very simple result, \eq{av}, appears again when one follows Pierre Curie\cite{Curie} and Pierre Weiss\cite{Weiss} in their development of a simplified theory of ferromagnetism.  Translated to the Ising case, their theory  would ask us to concentrate our attention upon one Ising variable, say the one at $\bf{r}$.   We would then notice that this one spin sees a Hamiltonian defined by
\be
- H_{{\bf r}}/T= \sigma_{\bf r} [ h({\bf r}) + K  \sum_{{\bf s}~nn~to~{\bf r}} \sigma_{\bf s}] +\text{~constant}
\la{environ}
\ee
where $h({\bf r})$ is the dimensionless magnetic field at ${\bf r}$ and the sum covers all the spins with positions, ${\bf s}$, sitting at nearest neighbor sites to ${\bf r}$. The remaining term, independent of $\sigma_{\bf r}$, has no direct effect upon the spin at $\bf{r}$.

\eq{environ} makes a statement about the environment faced by the statistical variable, $\sigma_{\bf r}$. At any given moment, it feels forces  that come from both the applied field and also the couplings to the spins in its immediate neighborhood. In the mean field theory these two kinds of fields are simply added.  However, in a real material, the effects of the field and the neighboring spins differ. The former is presumed to be time-independent while, in any real ferromagnet,  the latter will fluctuate in time.  For a given force-strength, one should expect a time-independent force to be more effectual in lining up spins and the fluctuating force to be less so.      Nonetheless, to make the spin problem tractable  Weiss\cite{Weiss}   made the approximation of ignoring this distinction. He replaced the actual, fluctuating values of the neighboring variables by their statistical averages.  Then the spin at ${\bf r}$ obeys exactly the same equation for its average value as the equation  for a single spin (see \eq{av}), except that the actual magnetic field is replaced by an effective field
\bsubs   \la{mft}
\be
h^{\text{eff}}({\bf r})=h({\bf r}) +K \sum_{{\bf s}~nn~to~ {\bf r}} < \sigma_{\bf s} >
\la{eff}
\ee
Now we have a very specific equation for the average spin at ${\bf r}$, namely
\be
<\sigma_{\bf r}> = \tanh h^{\text{eff}}({\bf r}) 
\la{avmft}
\ee
\esubs
Note that \eq{avmft} is of exactly the same form as \eq{av}.  The only difference between them is 
 the extra term in the equation for $h^{\text{eff}}$.

We can further reduce the complexity of these equations.  Let the applied field $h({\bf r})$ be independent of ${\bf r}$.  Then, if the system is large enough so that boundary effects do not matter,  no quantities in \eq{mft} will depend upon position and these equations may be simplified to
\be \la{Umft}
<\sigma>= \tanh h^{\text{eff}}    \text{~  where  ~ }h^{\text{eff}} = h +Kz < \sigma >
\ee
Here $z$ is the number of nearest neighbors of a given site.  For a simple cubic lattice in $d$-dimensions this number is given by $z=2d$.   In our further work, we shall write 
\be
T_c/T= Kz
\la{Tc}
\ee
since $K$ is inversely proportional to the temperature. As we shall see, $T_c$ is a critical temperature for this model. 

\begin{figure}
\begin{centering}
\includegraphics[height=7cm, ]{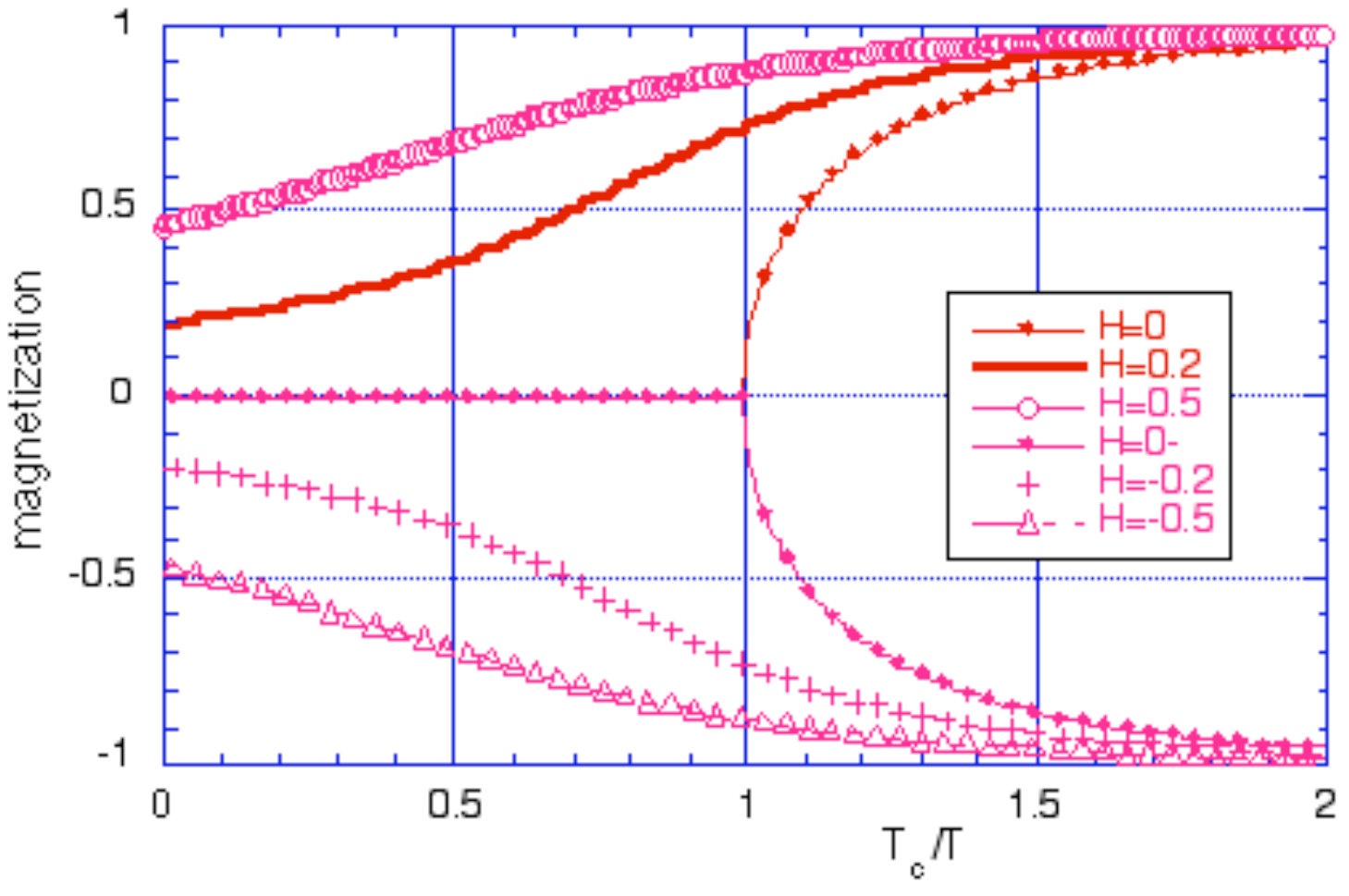}
\end{centering}
\caption{Magnetization, $<\sigma>$, versus temperature, $1/K$ for fixed values of the dimensionless magnetic field, here denoted as $H$, in replacement for what the text would have called $h$.  This graph was calculated from \eq{Umft} and \eq{Tc}.  }
\la{magnetization}
\end{figure}

\fig{magnetization} shows the form of the solution of \eq{Umft}.   The magnetization in this figure is simply defined to be $<\sigma>$, which is a dimensionless version of the magnetization density.   In the real world, the magnetization density would be this   average times the number of spins per unit volume times the  maximum magnetization per spin.  This figure defines curves of dimensionless magnetization versus temperature for various values of the dimensionless magnetic field, $h$. At high temperatures, as the magnetic field is varied,  the magnetization will vary continuously, tracking the magnetic field. Positive $h$ will give positive $<\sigma>$, negative $h$ will give negative $<\sigma>$, and $h$ equal to zero will give $<\sigma>=0$.  Next follow the curves for positive $h$ down to lower temperatures.  As the temperature is reduced, the magnetization will grow because the spins will tend to work together to align each other.  Equally, for negative $h$ as the temperature is reduced the magnetization becomes more negative. For $h=0$, the result is quite interesting.   For high temperatures, i.e. the left hand side of the graph, the $h=0$ magnetization remains zero at low temperatures.  At a special temperature, $T=T_c$ the $h=0$ curve splits into two branches.  (Physically, this split reflects a dichotomy in a real system that will spontaneously pick one of the two possible directions and line up the magnetization in that direction.  Usually there is some small field present and the system uses that field to ``decide'' which way to point.)   We see this splitting\footnote{Note that this mean field argument would equally well give a phase transition in a small system with periodic boundary conditions.   As Kramers has pointed out\cite{Uhlenbeck}, an infinite system is required to achieve a phase transition.  Consequently,  this apparent prediction of a phase transition in a finite system must be viewed as a grave failure of mean field theory.  This wrong prediction is a hint that mean field theory should fail in the neighborhood of the 
phase transition. }  as the system falling upon one or the other of the two curves shown for $h=0$.  Mathematically, there is a third solution to the mean field equations.  That solution has $<\sigma>=0$.  As discussed below, this third solution is not stable.  We need not follow it.     As the temperature is lowered still further the magnetization follows one of the two stable branches, as shown.  At the lowest temperatures, for all possible values of the applied field $h$ the magnetization falls into one of the two possible extreme values $<\sigma>=1$ or $<\sigma>=-1$.

How do we know that the system will not follow the solution with $h=<\sigma>=0$? This solution is ruled out by a thermodynamic stability argument originally used by James Clerk Maxwell\cite{Maxwell} in his discussions of fluid phase transitions.   In this magnetic case, the argument is very simple.  For fixed external parameters (like $h$ and $K$) the system always acts to adjust its internal parameters (here $<\sigma>$ ) to minimize the free energy.   A brief calculation compares the zero $<\sigma>$ with the other two solutions and gives the first of these a higher free energy.  The other two have equal free energies and can equally well exist at a given set of values of temperature and field.     

The phase diagram of \fig{ferro} follows from the calculation of these solutions to \eq{Umft}.         

The quantity $h^{\text{eff}}$ is termed a ``mean'' or ``effective'' field for this Ising problem.  Very many physics and chemistry problems have been solved by inferring a mean or effective field for the system and then using that field in a simple equation, usually one appropriate for a single particle system.   Often this approximation is used, as it is above, to describe a situation in which only a few other particles act upon the one in question. In these situations, the mean field theory approximation is only qualitatively accurate because the field in question is rapidly varying in time so that its average give a poor representation of its actual effect.  Nonetheless qualitative accuracy is often good enough for many useful applications.  

This kind of approximation might have seemed natural because it is exactly what we do in the theory of the electrodynamics of materials.  We describe the electrodynamics using four fields {\bf  E,B,D,H} two of which, ({\bf  D,H}),  are the field produced by effects external to the material, while the other two, ({\bf  E,B}), account for the averaged effects of charged particles within the material.  Since there are so very many charged particles in motion and since their effect is felt over a very long distance,   this use of effective fields can be quite accurate, at least in the classical applications involving large pieces of matter.\footnote{In recent applications to nanomaterials, this kind of approximation is often insufficient for a good description of the behavior of the materials. The averaging is inaccurate because the piece of material is too small.}  The considerable accuracy of mean field techniques in electrodynamics backs up the sometimes   less accurate use of mean field techniques throughout physics and chemistry.

\subsection{Meaning of the models}
\eqs{mft} define the mean field theory for an Ising model  magnet.  In fact, when there is a phase transition, these equations give a qualitatively correct description of what happens.   The first order phase transition happens at  low temperatures, which helps the system to line up its spins, even in the absence of a magnetic field.  The positive value of the coupling, $K$, means that two neighboring spins will tend to line up with one another.  Each of those spins will help align its neighbors. At two or higher dimensions, this  lining up will spread through the material through multiple chains of nearest neighbor alignments, which, all together, produce an overall alignment in the material.  These multiple chains permit the  spins to  line up in one particular direction even though there might be no external magnetic field tending to align them.  In the jargon of the field, they have attained their alignment ``self-consistently''. 

In this regard, it is quite telling that one-dimensional systems cannot show this alignment. They do have chains of alignment, just as in higher dimensions.  However, thermal fluctuations will, at any temperature above absolute zero, break the chains at some points.  Any breaks will cut the connections and thus will ruin the global alignment of the system.  Hence one dimensional systems show no ordered phases.  The mean field theory for these systems predicts ordering.  In this regard, mean field theory is entirely wrong.  On the other hand, in higher dimensions, one can imagine that multiple chains of nearest neighbor alignments connect any two far away points.  The chains have crosswise connections to keep the long-range alignment consistent even though fluctuations ruin some of the local alignments. These mutually reinforcing local alignments can thus reach out and result in global correlations, in fact correlations over conceptually infinite distances.   Thus the first order phase transition in the Ising model, and indeed in real materials, is a result of local couplings transferring information about the local phase of the system even to its farthest reaches.

I should describe how we know that a system does indeed show phase transitions. I have already mentioned two sources of information in this regard: a discernment of qualitative differences among different phases of matter, as in \fig{iceberg} and a mathematical proof by Peierls and Griffiths.

There is another way of knowing that the Ising model will have a phase transition.     One can define a variety of dynamical models with the property that, if a finite system is run over a very long period of time, the model will reproduce the result of Gibbsian statistical mechanics.   In running these models,  if one is sufficiently patient the smaller systems will explore all the possible configurations and the average value of each of the spins will be zero. Thus, the finite system will not have a phase transition.   As the size of the system gets larger, the system will tend to get stuck and explore only a limited subset of the possible configurations. The dependence of the configurations explored upon the size, shape, and couplings within the system is a major subject of contemporary, Twenty-First Century, exploration.   I cannot fully review these dynamical investigations here.  Nor can I discuss the analogous problems and results that arise in the experimental domain.  I shall however give a tiny, superficial, overview. 

In the case in which the system under study is large in only one of its directions and much smaller in its other dimensions, the system is said to be ``one-dimensional''.  In that case, as predicted by Landau, there tends to be a rather full examination of the phase space.    In the case in which the system extends over a long distance in more than one of the possible directions, then the system can easily get stuck and explore only a limited portion of the available configuration space.   However, "getting stuck" is not a simple easily understood event.   There are at least three qualitatively different scenarios for a less than full exploration of the available configurations.   In one case, independent of the starting state the entire Ising system will go into one of two possible ``basins of attraction'' defined by the two possible directions of the magnetization.  This first scenario is the one described here, and closely follows the possible  behavior of some real  materials.  In a second scenario, it is possible for the system to find itself in one of many different regions in ``configuration space'' and only explore that relatively small region, at least in any reasonable period of time. Over longer times, the size of the region will grow very slowly, but the region never encompasses most of the possible configurations. The repeated exploration of a slowly growing region  of configurations is characteristic of a behavior described as ``glassy''.  Such glasses tend to occur in many materials with relatively strong interactions. They are believed to be in some cases a dynamical property of materials, and in others an equilibrium property described by an extension of  Gibbsian statistical mechanics.   Present-day condensed matter science does not understand glassy behavior. A third scenario has the system divide in a  time-independent fashion into different regions, called {\em domains} or {\em grains}, each with its own phase.     

Still another way of knowing about phase transitions comes from the exact solutions of very simple models.  The first such solution was a calculation of the $h=0$ properties of the two-dimensional Ising model due to Lars Onsager\cite{Onsager}, followed by a calculation of the zero field magnetization by C. N. Yang\cite{Yang}.   These calculations show behavior somewhat similar to that described by mean field theory, but quite different in significant details, particularly near the critical point.  I hope to come back to this point is a later publication.

\subsection{Johannes  van der Waals and the theory of fluids}
Pierre Curie based his understanding of ferromagnets\cite{Curie} in part upon the earlier (1873)
van der Waals\cite{Waals} theory of the behavior of liquids.  Van der Waals started from the known relation between the pressure and the volume of a perfect gas, i.e. one that has no interactions between the molecules.  Expressed in modern form, the relation is  
\be
p=  T~N /V
\la{perfect}
\ee
Here, $p$ is the pressure, $V$ is the volume of the container, $N$ is the number of molecules within it  and $T$ is the temperature expressed in energy units.  This {\em equation of state} relates the pressure, temperature, and density of a gas in the dilute-gas region in which we may presume that interactions among the atoms are quite unimportant.   It says that the pressure is proportional to the density of particles, $N/V$, and to the temperature,  $T$.         This result is inferred by ascribing an average kinetic energy to each molecule proportional to $T$ and then calculating the transfer of momentum per unit area to the walls.  The pressure is this transfer per unit time.  Of course,  \eq{perfect} does not allow for any phase transitions.

Two corrections to this law were introduced by van der Waals to  estimate, in an approximate fashion, the effects of the interactions among the molecules. upon the properties of the fluid.   He intended to thereby account for the observed  phase diagram for a fluid. 

First, he argued that the molecules could not approach each other too closely because of an inferred short-ranged repulsive interaction among the molecules. This effect should reduce the volume available to the molecules by an amount proportional to the number of molecules in the system.  Thus, $V$ in \eq{perfect} should be replaced by the available or effective  volume, $V-Nb$, where $b$ would be the excluded volume around each molecule of the gas.  

The second effect is more subtle.    The pressure, $p$, is a force per unit area produced by the molecules hitting the walls of the container.  However, van der Waals inferred that there was an attractive interaction pulling each molecule towards its neighbors. This attraction is the fundamental reason why a drop of liquid can hold together and form an almost spherical shape.   As the molecules move toward the walls they are pulled back by the molecules they have left behind them, and their velocity is reduced. Because of this reduced velocity, their impacts impart less momentum to the walls    The equation of state contains the pressure as measured at the wall, $p$.  This pressure is the one produced by molecular motion inside  the liquid, $NT/(V-Nb)$, minus  the correction term coming from the interaction between the molecules near the walls.   That correction term is proportional to the density of molecules squared. In symbols the correction is $a(N/V)^2$  where $a$ is proportional to the strength of the interaction between molecules.      Van der Waals'  corrected expression for the pressure is thus
\be
p= NkT/(V-Nb)-a(N/V)^2
\la{Waals}
\ee
Here, $a$ and $b$ are parameters that are different for different fluids. 

\eq{Waals} is the widely used van der Waals equation of state for a fluid.  It is essentially a mean field equation, like the one, \eq{Umft}, that we used for the Ising model.  This equation of state can be used to calculate the particle density, $N/V$, as a function of temperature and pressure.  It is a cubic equation for $N/V$ and  has at most  three real solutions, rather like the ones we discussed earlier for $<\sigma>$.

The van der Waals equation of state provides a ``universal'' description that can be used to generate the phase diagram of a wide variety of liquids.  It is universal in the sense that if you make use of  the right variables, you will have an equation of state that applies equally well to most simple fluids.   One set of {\em right variables} are $p_R$, the pressure divided by the critical pressure,   $T_R$, the temperature divided by the critical temperature, and $n_R$, the number density (N/V) divided by its critical value.  In terms of these {\em reduced variables}, the van der Waals equation of state is\cite{Wiki1}       
\be
(p_R+3n_R^2)(\frac{1}{n_R}-\frac13)= 8T_R/3
\la{WaalsR}
\ee
This result can be used for practical purposes.  Using data from higher temperatures and pressures, Heike Kamerlingh Onnes\cite{Sengers} was able to extrapolate the low-temperature properties of helium fluid.  In this way he gained enough information to design the first-time-ever liquefaction of helium gas.       

\begin{figure}
\begin{centering}
\includegraphics[height=9cm ]{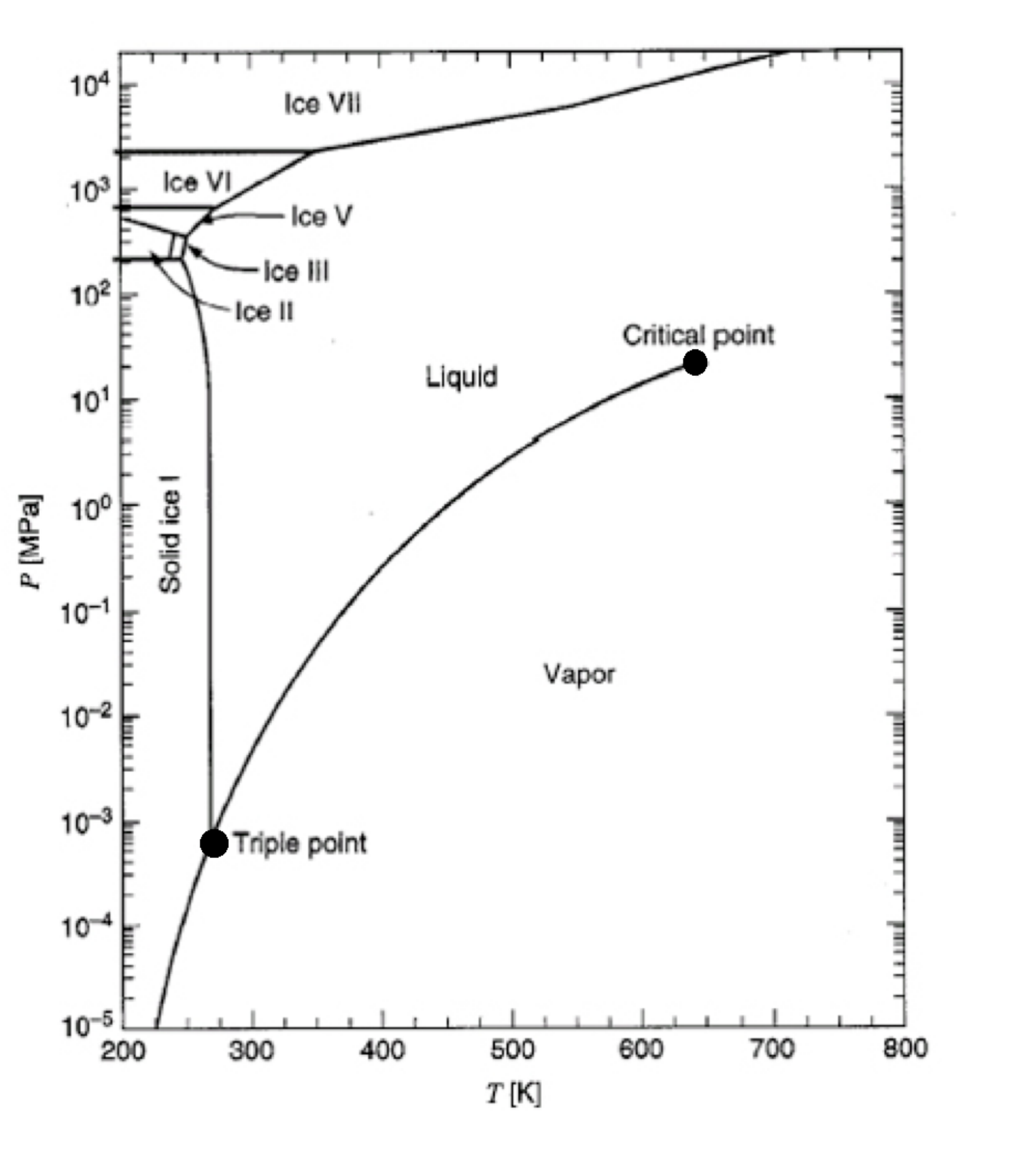}
\end{centering}
\caption{Phase Diagram for Water.  The curved line drawn between the triple point and the critical point is a locus of first order phase transitions as are all the lines delimiting  solid (Ice) phases. The critical point is a continuous phase transition.  Notice the possibility of progressing from the liquid to the vapor phase without passing through a phase transition.     }
\la{PhaseDiagramWater}
\end{figure}

Using the van der Waals equation is a little trickier than the use of the corresponding equations for magnetic systems.   In the latter case, when we solved for the magnetization as a function of magnetic field and temperature, we knew exactly where the discontinuous jump would occur.    It was at zero magnetic field.  This simplification was possible because the system displayed an exact symmetry under the change of sign of the magnetization and the magnetic field.   In this fluid system, there is also a jump.  Here the jump is a discontinuous change in the density at fixed temperature and pressure.  The jump is, in fact, the familiar first order phase transition between a liquid (high density) and a vapor (low density).  It is the change from boiling water to the steam above it.  The existence of the jump might have been clear to van der Waals because his equations have a thermodynamic instability in the region of the jump.  Furthermore, experiments\cite{Andrews}-- or even a simple observation of boiling water-- make a jump quite obvious.  However, a superficial examination of the equation of state does not determine exactly where the jump occurs.    Calculating its position involved a little tricky thermodynamics-- provided, as we have said, by Maxwell\cite{Maxwell}.  The phase diagram for water is shown in \fig{PhaseDiagramWater}.   Except for the richness of the solid phases, it looks like a distorted version of the magnetic phase diagram.  Many workers, starting with Curie\cite{Curie} and Weiss\cite{Weiss}, have used this similarity. 

Philosophers of science will look at Maxwell's application of thermodynamics with some interest.    The philosophy literature contains a considerable discussion of Ernest Nagel's principle of reduction\cite{Nagel}  that describes how a more fundamental theory {\em reduces  } a less fundamental one in an appropriate limit\cite{Berry,Batterman}.  This will occur when the ideas and laws of the reducing theory implies all the ideas and laws of the reduced theory\cite{Batterman2}.\footnote{A physicist would use the word reduce differently, with the arrow going in the other direction. A physicist might say:  {\em Special relativity reduces to Galilean relativity when all speeds are small in comparison to the velocity of light.} This difference in usage has been the source of some confusion.}  An example often employed is that thermodynamics might be a reduction of statistical mechanics,  in the philosophers' sense.  However, here Maxwell extended a statistical calculation by using thermodynamics.  I would worry about whether there is a simple process of reduction at work between statistical mechanics and thermodynamics, or perhaps between any two parts of science.  One hint at possible complication in this case is the title of Erwin Schr\"{o}dinger's classic text  ``Statistical Thermodynamics''\cite{Schr}.  Similar titles have been used by many other authors.

The van der Waals equation of state is not very accurate.  This fact was known to van der Waals, who was particularly concerned that the model did not fit the experimental facts near the critical point.    The next section outlines the near-critical behavior for the Curie-Weiss model for the ferromagnet.  The ferromagnetic model's behavior is quite similar to that of the fluid, but simpler.

\subsection{Near-critical behavior of Curie-Weiss\cite{Curie,Weiss} model}

We write down once again the equations for the mean field theory of the Ising model, \eq{Umft}, in a form appropriate for the study of the equations in the critical region.  To do this, we keep only the linear term in $h$ and the deviation, $t=1-T_c/T$, from the critical temperature.  Since we expect multiple solutions for the average magnetization we keep the lowest non-linear terms in $<\sigma_{\bf r}>$.    We also assume that this quantity is slowly varying in space and hold on to the lowest spatial gradient term.   In this way we derive

\begin{equation}
t <\sigma> = h - \frac{1}{3} <\sigma>^3 +\nabla^2 <\sigma> /z
\label{10simp}
\end{equation}

One can get quite interesting results by studying this equation in the region in which two or more of its terms are of the same order of magnitude and the remaining terms are much smaller.    
  
Focus on the spontaneous magnetization.  Set $h=0$ in equation (\ref{10simp}) and neglect the spatial variation to find 
\begin{equation}
t <\sigma> = -\frac{1}{3} <\sigma>^3
\label{10spont}
\end{equation}
One solution is $<\sigma>=0$.  But since we have a cubic equation, 
there are three solutions.  By dividing out the common factor of $<\sigma>$, we find that 
\begin{equation}
t = -\frac{1}{3} <\sigma>^2
\label{10fact}
\end{equation}
For $T > T_c$ (that is $t > 0)$ this equation has no real solutions.  So the only
possibility is that the magnetization is zero when the field is zero.   Thus we have a
\index{disordered state}disordered state for $T>T_c$.  However, for $T<T_c$,
$t<0$,  equation (\ref{10fact}) has two solutions: 
\begin{equation}
<\sigma> = \pm \sqrt{-3t},
\label{10solns}
\end{equation}
showing that there are two different states of spontaneous magnetization. 
In each one, as
$T$ goes to $T_c$, the magnitude of the magnetization goes to zero as $\sqrt{T_c-T}$.

\subsection{Critical indices}

\eq{10solns} is a crucial result from the Curie-Weiss mean field theory. It describes in quantitative form the jump near criticality. J. H. van der Waals knew that there was indeed   a jump in the density in his own theory of the  liquid-gas phase transition,  After the work of Maxwell,  he  knew too that his  theory gave a jump proportional to  $\sqrt{T_c-T}$ as criticality was approached.  But, in addition, he knew  the experimental data of Thomas Andrews\cite{Andrews}.  As pointed out in detail by Levelt Sengers\cite{Sengers} in her excellent exposition, van der Waals knew that a fit of the form in \eq{10solns} did not work but that a closely related fit,
\be
\text{jump = constant} (-t)^\b  
\la{beta}
\ee
worked quite well if one picked $\b$ to be one third.   Many results for behavior near critical points can be expressed as powers laws like \eq{beta},   with exponents in them called {\em critical indices}.    They are very useful for characterizing the behavior near critical points. 

Despite van der Waals' concern about the discrepancy between mean field theory and experiment in the region of the critical point, hardly anyone focused upon this issue in the years in which mean field theory was first being developed.   There was no theory or model that yielded \eq{beta} with any power different from one half,  so there was no focus for anyone's discontent.  The situation in which an old point of view continues  on despite evidence to the contrary is exactly of the sort described by Thomas Kuhn\cite{Kuhn}.

\section{Mean Field Theory Generalized}

\subsection{Many different mean field theories}

Following upon the work of van der Waals and Weiss, a wide variety of systems were described by mean field theories. The theory of phase transitions involving the unmixing of fluids was developed by van der Waals himself\cite{Sengers}, while  such unmixing in solids was described by W. L. Bragg and E. J. Williams\cite{Bragg}.  Literally dozens of such theories were defined, culminating in the theory of superconductivity of Bardeen, Cooper and Schrieffer\cite{BCS}.  These theories are all 
 different in that they have different physical quantities playing the roles we have given to the
magnetic field, or $T-T_c$, or serving as the    order parameters.   The order parameter is a
quantity that undergoes a discontinuous jump in the first order transition, and hence labels the different states that can arise under the
same physical conditions. In the examples we have given, the magnetization is the order parameter of the ferromagnet, the density\footnote{More properly, this order parameter is the density minus the value of the density at the critical point, in symbols $\rho-\rho_c$.} is the order parameter in the liquid-gas transition. Much effort and ingenuity has gone into the discovery and description of the order parameter in other phase transitions.  In the anti-ferromagnetic transition\cite{neel} the order parameter is a magnetization that points in opposite directions upon alternating lattice sites. In ferroelectrics, it is the electric field within the material. The superfluid\cite{Griffin} and superconducting transition have as their order parameter the quantum wave function for a macroscopically occupied state.  Liquid crystals have order parameters reflecting possible different kinds of orientation of the molecules within a liquid.    The description of these different manifestations of the phase transition concept reflect more than a century of work in condensed matter physics, physical chemistry, and the areas labeled, e.g.,  as ceramics, metallurgy,  materials science, ...

\subsection{Landau's generalization}
Lev Landau followed van der Waals, Pierre Curie, and Ehrenfest  in noticing a deep connection among different phase transition problems\cite{Daugherty}.   Landau was the first to translate this observation into a mathematical theory.    Starting from the recognition that each phase  
transition was a manifestation of a broken symmetry,  he used the order parameter to describe the nature and  the extent of  symmetry breaking\cite{Landau}.

Landau generalized the work of others by writing the free energy as an integral over all space of an appropriate function of the order parameter. The dependence upon ${\bf r}$ indicates that the order parameter is considered to be a function of position within the system.  In the simplest case, described above, the phase transition is one in which the order parameter, say the magnetization, changes sign.\footnote{The symmetry of the phase transition is reflected in the nature of the order parameter, whether it be a simple number (the case discussed here), a complex number (superconductivity and superfluidity), a vector (magnetism), or something else.}  In that case, the appropriate free energy takes the form
\be
F= \int ~ d{\bf r} \big[ A + B h({\bf r}) \Psi({\bf r})  +C~ \Psi({\bf r})^2 + D ~\Psi({\bf r})^4 +  E~[\nabla \Psi({\bf r}) ]^2 +\cdots       \big]
\la{F}
\ee
where $A,B,C, \ldots $ are parameters that describe the particular material and $\Psi({\bf r})$ is the order parameter at spatial position ${\bf r}$.     In recognition of the delicacy of the critical point, each term containing $\Psi$ is considered to go to zero more rapidly than $\Psi({\bf r})^2$ as criticality is approached. 

Notice that \eq{F} contains no term cubic in the order parameter.  Often this term is ruled out by symmetry considerations.  When such a term is present, the Landau theory predicts that there will be no critical point.   

The next step is to use the well-known rule of thermodynamics that the free energy is minimized  by the  achieved value of every possible macroscopic thermodynamic variable within the system.   Landau made the bold step of taking the magnetization density at each point to be a thermodynamic variable that could be used to minimize the free energy.   Using the calculus of variations one then gets an equation for the order parameter: 
\be
0=  h({\bf r}) +(C/B)~ \Psi({\bf r})  +4  (D/B) ~\Psi({\bf r})^3+2  (E/B)~\nabla^2 \Psi({\bf r}) 
\la{deltaF}
\ee
Notice that this result is precisely of the same form as \eq{10simp}, our previous result for the mean field theory magnetization equation,  as applied near the critical point.     The $C$-term is identified by this comparison as being proportional to the temperature deviation from criticality, $C= B t/2$.
\begin{figure}
\begin{multicols}{2}
\includegraphics[height=7cm ]{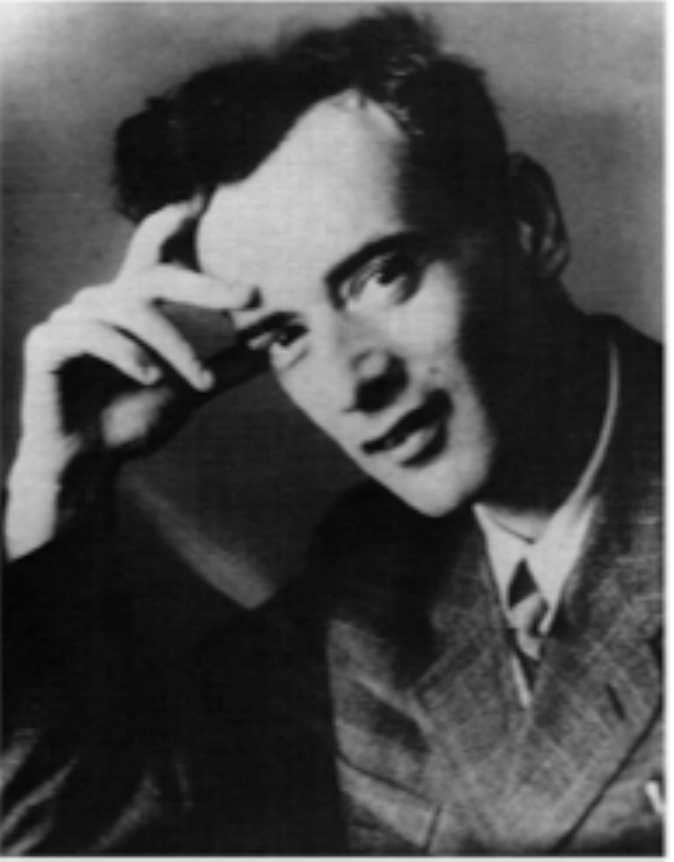}
\caption{L.D. Landau, 1908-1968.}
\la{landau}
\end{multicols}
\end{figure}

In some sense, of course, Landau's critical point theory is nothing new.  All his results are contained within the earlier theories of the individual phase transitions.   However, in another sense, his work was a very big step forward.   By using a single formulation that could encompass all critical phenomena with a given symmetry type, he pointed out the universality among different phase transition problems.   And indeed in the modern classification of phase transition problems\cite{67Review} the two main elements of the classification scheme are the symmetry of the order parameter and the dimension of the space.  Landau got the first one right but not, at least in this variational formulation, the second classifying feature.  In agreement with the current theory, the results do depend upon the symmetry properties of the free energy and the order parameter.  But the results disagree in that they do not depend upon the dimension of space.     On the other hand, Landau's inclusion of the space gradients seems, from a present-day perspective, to be right on.   In the current theory, the gradient terms tie together the  theory's  space dependence and its thermodynamic behavior.

There is also a deeply theoretical reason why his form for $F$  has provided a basis for most present-day work in critical phenomena and more broadly in statistical and particle physics.    His formulation of the problem via a free-energy and a variational statement,  ``F is minimized by a proper choice of $\Psi({\bf r})$,''  provides a clue to how one can approach these problems in a very powerful way.    In order to discuss modern problems in condensed matter physics, one must use the appropriate variable to describe the phenomenon  at hand.  One cannot   limit oneself to using just the variables that are handed down to us in conventional thermodynamics.   A formulation via a variational principle permits us to  use all the possible variables in the system to form an appropriate order parameter and to derive an equation for that constructed variable.    In this same vein, just a little later, and probably quite independently, both Julian Schwinger\cite{Milton},\cite{Galison} and Richard Feynman\cite{FH}  chose to use variational principles in their formulations of quantum physics and quantum field theory\cite{polaron}.  Their use were recognitions of the flexibility and depth of variational methods.

\subsection{And onward...}

My main story for this article ends in about 1937.   Here I look ahead to a few highlights of the succeeding years.  Following 1937, mean field theories were very extensively utilized to describe a very wide variety of phenomena and to reach more deeply  into what happens in a phase transition.  Work on the spatial variations within an ordered material,  which started with Ornstein and Zernike\cite{OZ} (who worked long before Landau) were carried forward using the gradient terms in the free energy by the Moscow school\cite{Landau,GL,Abrikosov}.  Wonderful extensions of mean field and effective field methods were brought  forward by Landau\cite{He3} and others.   I hope to tell some of this story in a subsequent publication.

Part of this additional story will focus upon critical behavior rather than the entire mean field phase diagram.  This behavior occurs in the neighborhood of the critical point, so that it is to be seen in only a very small region of the phase diagram of a typical system.  It is anomalous in that it is usually dominated by fluctuations rather than average values.  These two facts provide a partial explanation of why it took until the 1960s before it became a major scientific concern. Nonetheless, most of the ideas used in the eventual theoretical synthesis were generated before the Second World War.

The need for a new synthesis was emphasized by the workers at a conference on critical phenomena held at the U.S. National Bureau of Standards in 1965\cite{NBS}.  

 A new point of view had been called for by the work of the King's College school \cite{Domb,Fisher}, by the exact solution of the two-dimensional Ising model\cite{Onsager}, and, perhaps most important, by experimental observations\cite{Heller,67Review}.  Some development and integration of critical-point concepts were made in phenomenological work of the middle 1960s\cite{PP,Widom,LPK1966}.

Around 1970, these concepts were extended and combined with previous ideas from particle physics\cite{SP,GML} to produce a complete and beautiful theory of critical point behavior, the renormalization group theory of Kenneth G. Wilson\cite{Wilson}.
In the subsequent period this {\em revolutionary synthesis} radiated outward to (further) inform particle physics\cite{Cao},  nuclear physics, mathematical statistics,  and various dynamical theories.  It has become especially true that particle physics has absorbed the concepts that emerged from the deepened understanding of phase transitions.  The up and back relation exhibited here  between particle physics and condensed matter physics provides a counter-example to any one-way view of ``theory reduction''\cite{Nagel}. In one sense, particle physics ``reduces'' condensed matter physics. Nothing in condensed matter physics is likely to contradict quantum field theory.   But it is useful to note that, historically, in large part the flow of ideas in the physical sciences has been from the ``less fundamental'' subjects.  The down to earth physics observed in condensed matter and statistical physics has provided many of the ideas employed in the theory of particles and fields.

\section{Summary}
We have focused upon the nature of phase transitions and their description by the mean field theory originally set by van der Waals.  This theory provided a qualitative, but numerically inaccurate, description of the main events that occur in the phase diagrams of typical materials.  It explains the various ordering of these materials, and some of  their similarities and differences. It explains the universality of the phase diagrams, so that different materials can have rather similar phase diagrams. It fails to explain the lack of a phase transition in finite systems, or in systems that are infinite in only one dimension.  It fails to explain why the mean field theory phase diagrams are inaccurate near the critical point, exactly the region in which the general formulation of Landau would have them be most accurate.  Both mean field theory and Landau's variational method would be the main tools for the excellent advances to come.

\section*{Acknowledgment}
I had useful discussions related to this paper with Gloria Lubkin, E. G. D. Cohen, Michael Fisher, Roy Glauber, and Robert Batterman.    The material is this paper appeared in part in a talk at the 2009 seven pines meeting on the Philosophy of Physics under the title ``More is the Same,   
Less is the Same, too;  Mean Field Theories and Renormalization.''   This meeting was generously sponsored by Lee Gohlike.   This work was was supported in part by the University of Chicago MRSEC program under NSF grant number DMR0213745.  It was completed during a visit to the Perimeter Institute,  which is supported by the Government of Canada through Industry Canada and by the Province of Ontario through the Ministry of Research and Innovation.

\end{document}